\documentclass[]{jetpl}
\twocolumn
\usepackage{epsf}

\lat
\renewcommand{\vec}{\mathbf}

\title{Weak turbulence of gravity waves.}

\rtitle{Weak turbulence \ldots}

\sodtitle{Weak turbulence of gravity waves.}

\author{A.\,I.\,Dyachenko$^{+}$,
A.\,O.\,Korotkevich$^+$\/\thanks{e-mail: kao@landau.ac.ru}, V.\,E.\,Zakharov$^{+*}$}

\rauthor{A.\,I.\,Dyachenko, A.\,O.\,Korotkevich, V.\,E.\,Zakharov}

\sodauthor{Dyachenko, Korotkevich, Zakharov}

\address{$^+$L.D. Landau Institute for Theoretical Physics RAS,
119334 Moscow, Russia\\~\\
$^*$University of Arizona, Department of Mathematics, Tucson,  USA
}

\dates{16 April 2003}{*}

\abstract{For the first time weak turbulent theory was demonstrated for the surface gravity waves.
Direct numerical simulation of the dynamical equations shows Kolmogorov turbulent spectra
as predicted by analytical analysis \cite{Zakharov-DAN66} from kinetic equation.}


\begin{document}

\maketitle
In this Letter we study numerically the steady Kolmogorov spectra for spatially homogeneous gravity
waves. According to the theory of weak turbulence the main physical process here is the stationary
energy flow to the small scales, where the energy dissipates \cite{Zakharov-DAN66,Zakharov-JAMTP67}.
This flow is described by kinetic equation which has power-like solutions -- Kolmogorov spectra. This straightforward
picture takes place experimentally and numerically for different physical situations. For capillary waves it
was observed on the surface of liquid hydrogen \cite{Kolmakov-Lett}, \cite{Kolmakov}. The numerical
simulation of this process was performed in \cite{Pushkarev-96}. In nonlinear fiber optics these spectra
were demonstrated in numerical simulation \cite{Dyachenko-92}.There are many other results
\cite{Dias-2001,Vasiliev-2002,Lvov-2001,Galtier-2002,Musher-1995}. One of the most interesting
applications of the weak turbulence theory is the surface gravity waves. From the pioneering article
by Toba \cite{Toba} to the most recent observations \cite{Hwang} many experimentalists get the
spectra predicted by the weak turbulence theory. But these experiments cannot be treated as a complete
confirmation because the Zakharov-Filonenko spectrum is isotropic, while observed spectra are essentially
anisotropic. It is worth to say that the wave kinetic equation,
which is the keystone of this theory, was derived under several assumptions. Namely, it was assumed,
that the phases of all interacting waves are random and are in state of chaotic motion. The validity of
this proposition is not clear {\it a priori}. The direct numerical simulation of nonlinear dynamical equations
can give us a confirmation is this assumption valid or not. But for particular case of gravity surface waves
the numerical confirmation was absent in spite of significant efforts were applied. 
The only successful attempt in this direction was the simulation of freely
decaying waves \cite{Onorato-2002}. The reason for that for our opinion was concerned with a choice
of numerical scheme parameters. Namely, the numerical simulation is very sensitive to the width of
resonance of four-waves interaction. It must be wide enough to provide resonance on the discrete grid,
as it was studied in \cite{Capillary-2003} for decay of the monochromatic capillary wave. From the other
hand it has to be not too wide (due to nonlinear frequency shift) when the weak turbulent conditions fail.
We have spent significant efforts to secure the right choice of numerical parameters. As a result we have
obtained the first evidence of the weak turbulent Kolmogorov spectrum for energy flow for surface gravity
waves. The numerical simulation was surprisingly time consuming (in comparison to capillary waves
turbulence), but finally we clearly get spectrum for surface elevation
\begin{equation}
\left| \eta_k \right|^2 \sim \frac{1}{k^{7/2}},
\end{equation}
which is in the agreement with real experiments \cite{Toba,Hwang}.

{\it Theoretical background. --- }
Let us consider the potential flow of an ideal incompressible fluid
of infinite depth and with a free surface.
We use standard notations for velocity potential $\phi(\vec r, z, t),\vec r = (x,y); v= \nabla\phi$ and surface elevation
$\eta(\vec r, t)$. Fluid flow is irrotational $\triangle\phi = 0$. The total energy of the system can be
represented in the following form
$$
H = T + U,
$$
\begin{equation}
T = \frac{1}{2} \int d^2 r \int \limits_{-\infty}^{\eta} (\nabla \phi)^2 dz,
\end{equation}
\begin{equation}
U = \frac{1}{2} g \int \eta^2 d^2 r,
\end{equation}
where $g$ -- is the gravity acceleration. It was shown \cite{Zakharov-68} that under these assumptions
the fluid is a Hamiltonian system
\begin{equation}
\label{Hamiltonian_equations}
\frac{\partial \eta}{\partial t} = \frac{\delta H}{\delta \psi}, \;\;\;\;
\frac{\partial \psi}{\partial t} = - \frac{\delta H}{\delta \eta},
\end{equation}
where $\psi = \phi (\vec r, \eta (\vec r,t), t)$ is a velocity potential on the surface of the fluid. In order to
calculate the value of $\psi$ we have to solve the Laplas equation in the domain with varying
surface $\eta$. This problem is difficult. One can simplify the situation, using the expansion
of the Hamiltonian in powers of ''steepness''
\begin{equation}
\label{Hamiltonian}
\begin{array}{l}
\displaystyle
H = \frac{1}{2}\int\left( g \eta^2 + \psi \hat k  \psi \right) d^2 r + \\
\displaystyle
+ \frac{1}{2}\int\eta\left[ |\nabla \psi|^2 - (\hat k \psi)^2 \right] d^2 r + \\
\displaystyle
+ \frac{1}{2}\int\eta (\hat k \psi) \left[ \hat k (\eta (\hat k \psi)) + \eta\triangle\psi \right] d^2 r.
\end{array}
\end{equation}
For gravity waves it is enough to take into account terms up to the fourth order.
Here $\hat k$ is the linear operator corresponding to multiplying of Fourier harmonics by
modulus of the wavenumber $\vec k$.
In this case dynamical equations (\ref{Hamiltonian_equations}) acquire the following form
\begin{equation}
\label{eta_psi_system}
\begin{array}{lcl}
\displaystyle
\dot \eta &=& \hat k  \psi - (\nabla (\eta \nabla \psi)) - \hat k  [\eta \hat k  \psi] +\\
\displaystyle
		&&+ \hat k (\eta \hat k  [\eta \hat k  \psi]) + \frac{1}{2} \triangle [\eta^2 \hat k \psi] + 
		\frac{1}{2} \hat k [\eta^2 \triangle\psi],\\
\displaystyle
\dot \psi &=& - g\eta - \frac{1}{2}\left[ (\nabla \psi)^2 - (\hat k \psi)^2 \right] - \\
\displaystyle
		&& - [\hat k  \psi] \hat k  [\eta \hat k  \psi] - [\eta \hat k  \psi]\triangle\psi + D_{\vec r} + F_{\vec r}.
\end{array}
\end{equation}
Here $D_{\vec r}$ is some artificial damping term used to provide dissipation at small scales;
$F_{\vec r}$ is a pumping term corresponding to external force (having in mind wind blow, for example).
Let us introduce Fourier transform
$$
\psi_{\vec k} = \frac{1}{2\pi} \int \psi_{\vec r} e^{i {\vec k} {\vec r}} d^2 r,\;\;
\eta_{\vec k} = \frac{1}{2\pi} \int \eta_{\vec r} e^{i {\vec k} {\vec r}} d^2 r.
$$
With these variables the Hamiltonian  (\ref{Hamiltonian}) acquires the following form
\begin{equation}
\begin{array}{l}
\displaystyle
H = H_0 + H_1 + H_2 + ...,\\
\displaystyle
H_0 = \frac{1}{2}\int (|k| |\psi_{\vec k}|^2 + g |\eta_{\vec k}|^2)d\vec k,\\
\displaystyle
H_1 = -\frac{1}{4\pi}\int L_{\vec k_1\vec k_2 }\psi_{\vec k_1}\psi_{\vec k_2}\eta_{\vec k_3}\times\\
 \displaystyle
\times\delta (\vec k_1 + \vec k_2 + \vec k_3) d\vec k_1 d\vec k_2 d\vec k_3,\\
\displaystyle
H_2 = \frac{1}{16\pi^2}\int M_{\vec k_1\vec k_2\vec k_3\vec k_4}
\psi_{\vec k_1}\psi_{\vec k_2}\eta_{\vec k_3}\eta_{\vec k_4}\times\\
\displaystyle
\times\delta (\vec k_1 + \vec k_2 + \vec k_3 + \vec k_4)d\vec k_1 d\vec k_2 d\vec k_3 d\vec k_4,\\
\end{array}
\end{equation}
Here
\begin{equation}
\begin{array}{l}
\displaystyle
L_{\vec k_1 \vec k_2} = (\vec k_1 \vec k_2) + |k_1||k_2|,\\
\displaystyle
M_{\vec k_1\vec k_2\vec k_3 \vec k_4} = |\vec k_1| |\vec k_2|\left[
\frac{1}{2}(|\vec k_1 + \vec k_3| + |\vec k_1 + \vec k_4| + \right.\\
\displaystyle
\left. +|\vec k_2 + \vec k_3| + |\vec k_2 + \vec k_4|) -
|\vec k_1| - |\vec k_2|\right].\\
\end{array}
\end{equation}
It is convenient to introduce the canonical variables $a_{\vec k}$ as shown below
\begin{equation}
\label{a_k_substitution}
a_{\vec k} = \sqrt \frac{\omega_k}{2k} \eta_{\vec k} + i \sqrt \frac{k}{2\omega_k} \psi_{\vec k},
\end{equation}
where
\begin{equation}
\label{dispersion_relation}
\omega_k = \sqrt {g k},
\end{equation}
this is the dispersion relation for the case of infinite depth. The similar formulas can be derived in the
case of finite depth \cite{Zakharov-1999}.
With these variables the equations (\ref{Hamiltonian_equations}) take the following form
\begin{equation}
\label{Hamiltonian_eqs_canonical}
\dot a_{\vec k} = -i \frac{\delta H}{\delta a_{\vec k}^{*}}.
\end{equation}
The dispersion relation (\ref{dispersion_relation}) is of the ''non-decay type'' and the equations
\begin{equation}
\omega_{k_1} = \omega_{k_2} + \omega_{k_3},\;\;\; \vec k_1 = \vec k_2 + \vec k_3
\end{equation}
have no real solution. It means that in the limit of small nonlinearity, the cubic terms
in the Hamiltonian can be excluded by a proper canonical transformation
$a(\vec k,t) \longrightarrow b(\vec k,t)$ \cite{Springer-92}. The formula of this transformation
is rather bulky and well known \cite{Zakharov-1999,Springer-92}, so let us omit the details here.

For statistical description of a stochastic wave field one can use a pair correlation function
\begin{equation}
<a_{\vec k} a_{\vec k'}^*> = n_{k} \delta (\vec k - \vec k').
\end{equation}
The $n_{\vec k}$ is measurable quantity, connected directly with observable correlation functions.
For instance, from (\ref{a_k_substitution}) one can get
\begin{equation}
\label{I_k_expression}
I_k = <|\eta_{\vec k}|^2> = \frac{1}{2}\frac{\omega_k}{g} (n_k + n_{-k}).
\end{equation}
In the case of gravity waves it is convenient to use another correlation function
\begin{equation}
<b_{\vec k} b_{\vec k'}^*> = N_{k} \delta (\vec k - \vec k').
\end{equation}
The function $N_k$ cannot be measured directly. The relation connecting $n_k$ and $N_k$
is rather complex in the case of fluid of finite
depth. But in the case of deep water it becomes very simple \cite{Zakharov-1999}
\begin{equation}
\frac{n_k - N_k}{n_k} \simeq \mu,
\end{equation}
where $\mu = (ka)^2$, here $a$ is a characteristic elevation of the free surface. In the case of the weak
turbulence $\mu << 1$. The correlation function $N_k$ obey the kinetic equation \cite{Zakharov-DAN66}
\begin{equation}
\label{Kinetic_equation}
\frac{\partial N_k}{\partial t} = st(N,N,N) + f_p (k) - f_d (k),
\end{equation}
Here 
\begin{equation}
\begin{array}{l}
\displaystyle
st(N,N,N)=4\pi \int \left| T_{\vec k,\vec k_1,\vec k_2,\vec k_3}\right|^2 \times\\
\displaystyle
\times(N_{k_1} N_{k_2} N_{k_3} +N_{k} N_{k_2} N_{k_3} - N_{k} N_{k_1} N_{k_2} -\\
\displaystyle
- N_{k} N_{k_1} N_{k_3})\delta (\vec k + \vec k_1- \vec k_2 - \vec k_3)d \vec k_1d \vec k_2 d \vec k_3.\\
\end{array}
\end{equation}
The complete form of matrix element $T_{\vec k,\vec k_1,\vec k_2,\vec k_3}$ can be found in many sources
\cite{Zakharov-DAN66, Zakharov-JAMTP67,Zakharov-1999}.
Function $f_p (k)$ in (\ref{Kinetic_equation})
corresponds to wave pumping due to wind blow for example. Usually it is located on long scales. Function $f_d (k)$
represents the absorption of waves due to viscosity and wave-breaking. None of this functions are known to a sufficient
degree.

Let us consider stationary solutions of the equation (\ref{Kinetic_equation}) assuming that
\begin{itemize}
\item The medium is isotropic with respect to rotations;
\item Dispersion relation is a power-like function $\omega=a k^\alpha$;
\item $T_{\vec k,\vec k_1,\vec k_2,\vec k_3}$ is a homogeneous function:
$T_{\epsilon \vec k,\epsilon \vec k_1,\epsilon \vec k_2,\epsilon \vec k_3} = \epsilon^\beta T_{\vec k,\vec k_1,\vec k_2,\vec k_3}$.
\end{itemize}
Under this assumptions one can get Kolmogorov solutions \cite{Springer-92}
\begin{equation}
\begin{array}{l}
\displaystyle
n_k^{(1)} = C_1 P^{1/3} k^{-\frac{2\beta}{3} - d},\\
\displaystyle
n_k^{(2)} = C_2 Q^{1/3} k^{-\frac{2\beta - \alpha}{3} - d}.\\
\end{array}
\end{equation}
Here $d$ is a spatial dimension ($d=2$ in our case).
The first one is a Kolmogorov spectrum, corresponding to a constant flux of energy $P$ to the region of small scales
(direct cascade of energy). The second one is Kolmogorov spectrum, describing inverse cascade of wave action to large scales,
and $Q$ is a flux of action. In both cases $C_1$ and $C_2$ are dimensionless ''Kolmogorov's constants''.

In the case of deep water $\omega=\sqrt{gk}$ and, apparently, $\beta=3$.
It is known since \cite{Zakharov-DAN66} that on deep water
\begin{equation}
n_k^{(1)} = C_1 P^{1/3} k^{-4}.
\end{equation}
In the same way \cite{Zakharov-1982} for second spectrum
\begin{equation}
n_k^{(2)} = C_2 Q^{1/3} k^{-23/6}.
\end{equation}

In this Letter we will explore the first spectrum (energy cascade). Using (\ref{I_k_expression}) one can get
\begin{equation}
I_k = \frac{C_1 g^{1/2} P^{1/3}}{k^{7/2}}.
\end{equation}

{\it Numerical Simulation ---} Dynamical equations (\ref{eta_psi_system}) are very hard for analytical analysis. One
of the main obstacles is the $\hat k $-operator which is nonlocal. However, using Fourier technique
practically makes no difference between derivative and $\hat k$. The numerical simulation
of the system is based upon consequent application of fast Fourier transform algorithm. The details of this
numerical scheme will be published separately.

For numerical integration of (\ref{eta_psi_system}) we used the functions $F$ and $D$ defined in Fourier
space
\begin{equation}
\begin{array}{l}
\displaystyle
F_k = f_k e^{iR_{\vec k} (t)},\\
\displaystyle
f_k = 4 F_0 \frac{(k-k_{p1})(k_{p2}-k)}{(k_{p2} - k_{p1})^2};\\
\displaystyle
D_{\vec k} = \gamma_k \psi_{\vec k},\\
\displaystyle
\gamma_{k} = -\gamma_1, k \le k_{p1},\\
\displaystyle
\gamma_{k} = - \gamma_2 (k - k_d)^2, 
k > k_d.
\end{array}
\end{equation}
Here $R_{\vec k} (t)$ is the uniformly distributed random number in the interval $(0,2\pi)$.
We have solved system of equations (\ref{eta_psi_system}) in the periodic
domain $2\pi\times2\pi$ (the wave-numbers $k_x$ and $k_y$ are integers in this case).
The size of the grid was chosen $256\times256$ points. Gravity acceleration $g=1$.
Parameters of the damping and pumping were the following: $k_{p1} = 5,\; k_{p2} = 10,\; k_d = 64$.
Thus the inertial interval is about half of decade.

During the simulations we paid special attention to the problems which could ''damage'' the calculations.
First of all, the ''bottle neck'' phenomenon
at the boundary between inertial interval and dissipation region.
This effect is very fast, but can be effectively suppressed by proper choice of damping value $\gamma_2$ in the
case of moderate pumping values $F_0$.
The second problem is the accumulation of ''condensate'' in low wave numbers.
This mechanism for the case of capillary waves
was examined in details in \cite{Capillary-2003}. This obstacle can be overcome by simple adaptive
damping scheme in the small wave numbers. After some time system reaches the stationary state,
where the equilibrium between pumping and damping takes place. Important parameter in this state
is the ratio of nonlinear energy to the linear one $(H_1 + H_2)/H_0$.

\begin{figure}[hbt]
\centerline{\epsfxsize=8.5cm \epsfbox{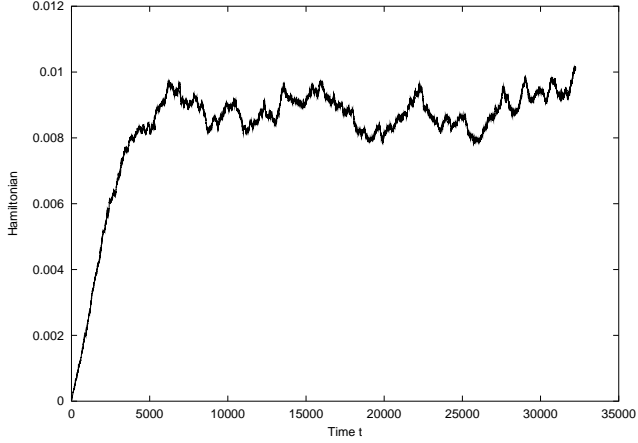}} 
\vspace{4mm}
\caption[]{\label{Hamiltonian_fig}Fig.1. Hamiltonian as a function of time.}
\end{figure}

For example, in the case of $F_0 = 2\times10^{-4}, \gamma_1 = 1\times10^{-3}, \gamma_2 = 400$
the level of nonlinearity was equal to
$(H_1 + H_2)/H_0 \simeq 2\times10^{-3}$. The Hamiltonian as a function of time is shown in
Fig. \ref{Hamiltonian_fig}.

\begin{figure}[hbt]
\centerline{\epsfxsize=8.5cm \epsfbox{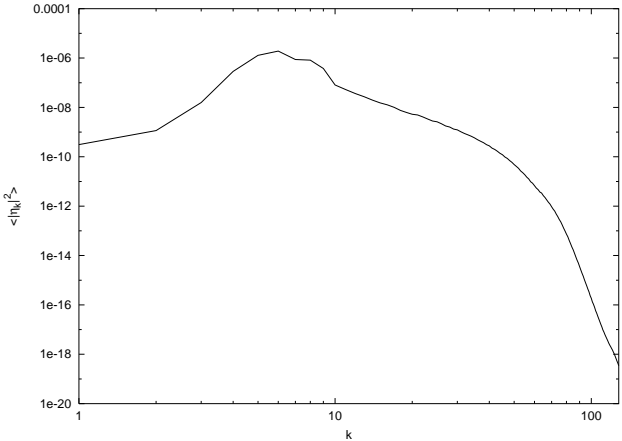}} 
\vspace{4mm}
\caption[]{\label{Correlator_fig}Fig.2. The logarithm of the correlator function of surface elevation
as a function of logarithm of the wave number.}
\end{figure}

The surface elevation correlator function appears to be power-like in the essential part of inertial interval,
where the influence of pumping and damping was small. The correlator is shown in Fig. \ref{Correlator_fig}.

\begin{figure}[hbt]
\centerline{\epsfxsize=8.5cm \epsfbox{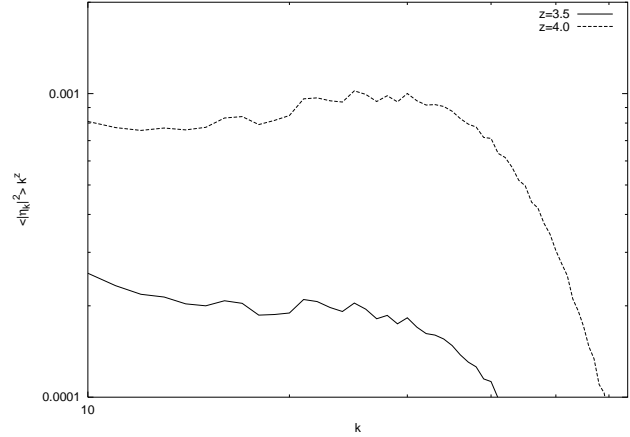}} 
\vspace{4mm}
\caption[]{\label{Correlator_mul}Fig.3. Compensated correlators in inertial interval for different values of the compensation
power: $z=3.5$ solid line (weak turbulence theory), $z=4.0$ dashed line (Phillips theory).}
\end{figure}

One can try to estimate the exponent of the spectrum. It is worth to say that an alternative spectrum
was proposed earlier by Phillips \cite{Phillips-1958}. That power-like spectrum is due to wave
breaking mechanism
and gives us a surface elevation correlator as $I_k \sim k^{-4}$.
Compensated spectra are shown in the Fig. \ref{Correlator_mul}. It seems to be an evidence,
that the Kolmogorov spectrum predicted by weak turbulence theory better fit the results of the
numerical experiment.

The inertial interval was rather narrow (half a decade). But the obtained results allow us to conclude,
that accuracy of experiment was good enough under the time constraints of simulation (we get the steady
state after 20-30 h using available hardware, and we need several days to average $|\eta_k|^2$ function).
The simulation on larger grid ($512\times512$, for example) can make the accuracy better. But even
these results give us a clear qualitative picture.

This work was
supported by RFBR grant 03-01-00289, INTAS grant 00-292, the Programme
``Nonlinear dynamics and solitons'' from the RAS Presidium and ``Leading Scientific
Schools of Russia" grant, also by US Army Corps of Engineers, RDT\&E Programm,
Grant DACA 42-00-C0044 and by NSF Grant NDMS0072803.

Also authors want to thank creators of the opensource fast Fourier transform library
FFTW \cite{FFTW} for this fast, portable and completely free piece of software.

\end{document}